\documentclass[aip,apl,amsmath,amssymb,reprint]{revtex4-1}
\usepackage{graphicx}% Include figure files
\usepackage{dcolumn}% Align table columns on decimal point
\usepackage{bm}
\usepackage{amsmath}
\usepackage{graphics}

\begin{document}

%\preprint{AIP/123-QED}
\title{Photon-Induced Selenium Vacancies in TiSe$_2$}

\author{David B. Lioi}
 \affiliation{Department of Physics, Drexel University, Philadelphia, Pennsylvania 19104, USA}

\author{David J. Gosztola}%
\author{Gary P. Wiederrecht}
\affiliation{%
 Center for Nanoscale Materials, Argonne National Laboratory, Argonne, Illinois 60439, USA}

\author{Goran Karapetrov}
\email{goran@drexel.edu}
\affiliation{Department of Physics, Drexel University, Philadelphia, Pennsylvania 19104, USA}

\date{\today}

\begin{abstract}
TiSe$_2$ is a member of transition metal dichalcogenide family of layered van-der-Waals materials that exhibits some distinctive electronic and optical properties. Here, we perform Raman spectroscopy studies on single crystal TiSe$_2$ to investigate photon-induced defects associated with formation of
selenium vacancies. Two additional E$_g$ phonon peaks are observed
in the laser-irradiated regions, consistent with Raman spectra of
selenium deficient TiSe$_2$. Temperature dependent studies of the
threshold laser intensity necessary to form selenium vacancies
show that there is a linear dependence. We extract the relevant
activation energy for selenium vacancy nucleation. The impact of
these results on the properties of strongly correlated
electron states in TiSe$_2$ are discussed.
\end{abstract}


\maketitle

%\section{Introduction} \label{intro}
Transition metal dichalcogenides (TMDs) exhibit
series of unique electronic properties ranging from charge density
wave (CDW) order to superconductivity. 1T-TiSe$_2$~\ is a quasi-2D
layered material with a trigonal symmetry that has been studied
for over 30
years~\cite{Salvo_1976_PRB,Wilson_1978_PSSB,Guo_2015_PRB,Lorchat_2016_ACSNano,Cercellier_2007_PRL,Monney_2010_NJP,Fogler_2014_NatureCom,Korn_2011_APL,Bhatt_2013_APMSP}.
Only recently it was discovered that the CDW in this material has
excitonic origin~\cite{Cercellier_2007_PRL,Monney_2010_NJP} and a chiral
order,~\cite{Ishioka_2010_PRL,Castellan_2013_PRL} which might have
implications on fundamental understanding of the strongly
correlated electron systems. On the other hand, the ability to
separate charges in materials with reduced dimensionality is of
great interest for many potential applications.~\cite{Fogler_2014_NatureCom,Korn_2011_APL}
Due to its specific structural and electronic properties TiSe$_2$
has also been considered as alternative to graphene electronics,~\cite{TCChiang_2015_NatComm,Li_2016_Nature} in thermoelectric
applications~\cite{Bhatt_2014_ACSAMI,Bhatt_2013_APMSP} and as a cathode material
in batteries.~\cite{Gu_2015_ScientificRep}

The strongly correlated electron behavior in TMDs is usually
attributed to their quasi-two-dimensional structure consisting of
$X$-$M$-$X$ layers ($M-$transition metal and $X-$chalcogen) weakly
bound together by van der Waals forces.~\cite{Lv_2015_ACR}
Further interest in these systems has been stimulated by the
changes in their bulk electronic properties upon reduction to a
single monolayer. The modification of the electronic properties in
2D layered systems is due to the changes in the electronic and
phonon band structure when the systems are reduced to a single
$X$-$M$-$X$ layer.~\cite{Zhu_2011_PRB,Lebegue_2009_PRB} Fabrication
of single or few layer TMDs proceeds either by exfoliation from
single crystals~\cite{Ulbaldini_2014_JoCG,Li_2016_Nature} or through
chemical growth of thin
films.~\cite{Peng_2015_PRB,Boscher_2006_CVD,Sugawara_2016_ACSNano} The
first method has been favored by many researchers ever since it
was first used with graphene, as it tends to yield better
crystalline quality and, therefore, is very suitable for studying
the fundamental physical properties. However, exfoliation is not
scalable, making it ill-suited for device fabrication and
integration with other thin film materials and processes.

On the other hand, physical and chemical vapor deposition methods
are very flexible in terms of control of thickness and chemical
composition of the dichalcogenide layers. Up to now TiSe$_2$
monolayers have proven difficult to obtain through methods other
than molecular beam epitaxy.~\cite{Sugawara_2016_ACSNano}  Synthesized
thin film TMDs often suffer from high concentration of defects and
lack of long range crystalline order~\cite{Lv_2015_ACR}. In
TiSe$_2$ the most frequent defects are selenium
vacancies~\cite{Salvo_1976_PRB,Hildebrand_2014_PRL}. The concentration
of selenium vacancies depends strongly on the conditions of film
growth, especially growth temperature, similar as in the case of
single crystal growth~\cite{Salvo_1976_PRB,Wilson_1978_PSSB}.
Selenium deficiencies are often detrimental to the charge density
wave order~\cite{Lv_2015_ACR} as their presence
changes the nature of the material from semiconducting to
semimetallic. Therefore, the characterization of the defects and
their dynamics is of utmost importance.

Raman spectroscopy is one of the main tools utilized to study
layered TMDs, as the symmetric A$_{1g}$ breathing phonon mode and
E$_g$ layer shearing  mode are sensitive to inter- and intra-layer
properties of the structure.~\cite{Zhao_2011_JACS} As in the case of
graphene, Raman spectra can distinguish between single and few
layers of TMD on the surface. Indeed, both resonant and
non-resonant Raman spectroscopy has already been used to
distinguish multilayered thin films of
MoS$_2$,~\cite{Soubelet_2016_PRB}
WS$_2$,~\cite{Staiger_2015_PRB} MoTe$_2$,~\cite{Guo_2015_PRB}
and other members of the TMD family.~\cite{Lorchat_2016_ACSNano} In the
present work we study the evolution of selenium vacancies in high
quality TiSe$_2$ single crystals as they appear in Raman spectra
through Raman laser induced vaporization of the selenium atoms.
The temperature dependent studies allow us to estimate the
activation energy for selenium vacancy nucleation, an important
parameter if considering TiSe$_2$ as material for electronic and
optical applications.

%\section{Results} \label{sec:1}
High quality single crystals of TiSe$_2$ were grown
using chemical vapor transport
method~\cite{Salvo_1976_PRB,Oglesby_1994_JoCG,Wilson_1978_PSSB} and
characterized using EDS, XRD and variable temperature electrical
transport. One of the signatures of low level of intrinsic defects in the
single crystals is the high charge density peak in resistivity at
the CDW transition (2.5--3.5 times the room temperature resistivity)
and low level of defect concentration as observed by scanning
tunneling microscopy~\cite{Iavarone_2012_PRB}. Variable temperature
Raman spectroscopy measurements were conducted at the Center for
Nanoscale Materials at Argonne National Laboratory using a
Renishaw InVia Raman microscope with a 514 nm argon ion laser
source and a $\sim$\,1.5\,$\mu$m diameter spot size. The spectrometer
is equipped with variable temperature cell capable of operating
between 80--400\,K. All the experiments were conducted in the
presence of ultra-high pure nitrogen exchange gas at normal
pressure.
\begin{figure}
    \includegraphics[width=8cm]{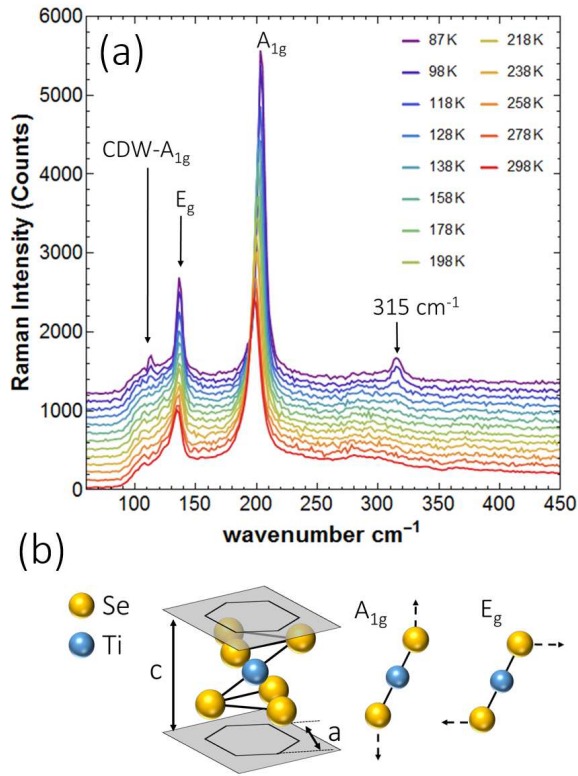}
    \caption{(a) Temperature evolution of Raman spectra of TiSe$_2$
        single crystal (1.05\,mW excitation power); (b) structure of the
        unit lattice of 1T-TiSe$_2$ and Raman active breathing mode
        (A$_{1g}$) and shear mode (E$_g$) of the crystal lattice.}
    \label{fig1}
\end{figure}
Fig.~\ref{fig1}a shows Raman spectra on freshly cleaved surface of
TiSe$_2$ single crystal. At room temperature one can see the
normal phase shearing mode E$_g$ and breathing mode A$_{1g}$
peaks. As the temperature is lowered below the CDW transition of
$\sim$\,200\,K we see the emergence of additional peaks that reflect
the change in lattice symmetry. The CDW doubles the period of the
crystal lattice to 2a$_0\times$2a$_0\times$2c and induces
additional Raman peaks: the 315\,cm$^{-1}$ peak is first seen below
$\sim$\,160\,K, in agreement with previous
measurements.~\cite{Snow_2003_PRL,Sugai_1980_SSC,Holy_1977_PRB} The charge density wave
A$_{1g}$ peak at 110\,cm$^{-1}$ becomes fully observable only at
temperatures lower than $\sim$\,100K.~\cite{Barath_2008_PRL}

\begin{figure}
    \includegraphics[width=8cm]{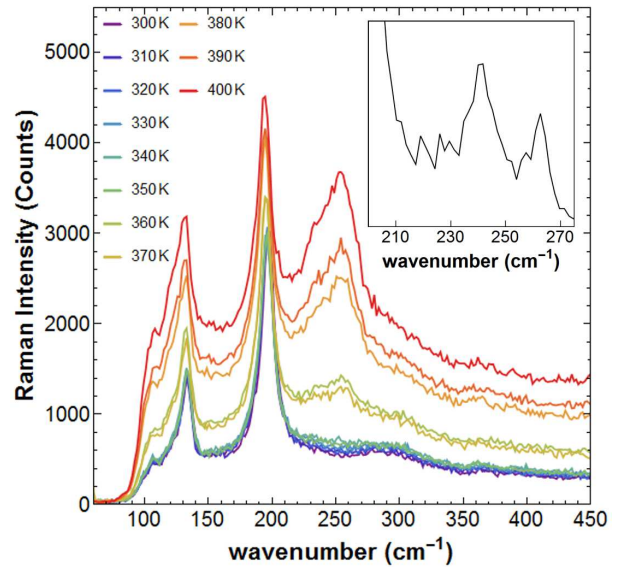}
    \caption{Raman spectra of TiSe$_2$ single crystal at different
        temperatures using 1.87mW laser excitation power ($\lambda=$~514\,nm,
        spot diameter is 1.5\,$\mu m$, averaging time is 3 minutes).
        Each spectrum is taken at pristine location on the freshly cleaved
        surface of TiSe$_2$ single crystal. Clear onset of the laser
        damage threshold is at around 355\,K. The inset shows expanded view of the
        spectrum obtained with low excitation power ($\sim$\,100\,$\mu$W) at
        a location irradiated at 400\,K} \label{fig2}
\end{figure}

Uniform heating of the TiSe$_2$ single crystal above room
temperature makes the crystal structure more susceptible to laser
induced change. Fig.~\ref{fig2} shows the evolution of Raman
spectra of the TiSe$_2$ single crystal from room temperature and
up to 400\,K. At 300\,K one notices a broad Raman background signal
centered at around 250\,cm$^{-1}$. There are also enhanced
fluctuations of the background intensity in the spectrum around
this frequency. As the the temperature is increased, the broad
background slowly evolves into a set of distinct peaks. At 360\,K we
observe a broad increase in the background at near 250\,cm$^{-1}$.
This broad peak is consistent with previous Raman experiments
performed on non-stoichiometric TiSe$_{2-x}$ with high
concentration of selenium vacancies.~\cite{Bhatt_2014_ACSAMI} As the
temperature is raised further, the broad background evolves into
distinct peaks that are made up of two Raman modes centered at 241
and 263\,cm$^{-1}$. They are shown in the inset of Fig.\ref{fig2}
and were obtained by reducing the laser power from 1.87\,mW to 1.05\,mW
and moving the laser spot outside the perimeter of the original
beam profile. The result indicates that high beam powers damage
the sample surface within the laser beam profile, and within a
much larger radius ($\sim$\,100\,$\mu$m depending on the excitation
power) some of the surface selenium atoms sublime leaving selenium
vacancies on the TiSe$_2$ single crystal surface.

\begin{figure}
    \includegraphics[width=8cm]{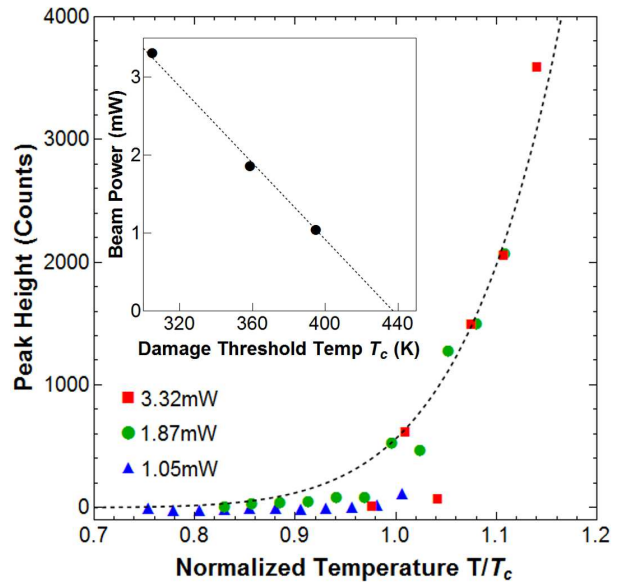}
    \caption{Raman E$_g$ peak intensity at various temperatures and
        laser powers. The temperature axis is normalized to the
        threshold temperature T$_c$ of laser induced selenium vacancy
        generation. The inset shows the dependence of the threshold
        temperature T$_c$ on laser beam power.  Exposure time of 3\,minutes
        was used for all experiments and each point was taken on
        unirradiated pristine area of the surface. The schematic
        illustrates the E$_g$ vibration mode in the vicinity of the
        selenium vacancy.}
    \label{fig3}       % Give a unique label
\end{figure}

To understand the relevant temperatures and activation energies
for nucleation of selenium vacancies we conducted systematic
experiments by varying both laser beam power and temperature. Each
measurement was made on large atomically flat planes of the single crystal
surface with no edges in the microscope image. Raman spectrum at
each temperature was taken for 3\,min before blocking the laser
beam and moving on to the next higher temperature setpoint. Each
spectrum was taken on a different and pristine location on the
surface of the single crystal in order to avoid effects from previous laser irradiation.
In Fig.~\ref{fig3} Raman intensity at 253\,cm$^{-1}$ is plotted
versus temperature for multiple beam powers, each showing
different onset of the peak increase with temperature. The
threshold temperature beyond which the peak height at 253\,cm$^{-1}$
starts increasing marks the onset of irreversible damage
done to the crystal surface. The inset in Fig.~\ref{fig3} shows a
linear relationship between the beam power and threshold
temperature T$_c$ for irreversible damage, suggesting that thermal
effects are responsible for selenium vaporization. The
extrapolation of the curve to vanishing laser power leads to zero
crossing at 435\,K. This suggest of a relatively low temperature
threshold at which the process of generation of selenium vacancies
starts taking place. Since this process is thermally activated, we
fit the peak height vs normalized temperature with Arrhenius
dependence and find that the activation energy for selenium
vacancy generation is ~0.52$\pm0.07$\,eV. Our results of laser
damage of TiSe$_2$ single crystals very much resemble the
experiments on CdZnTe in which the processes of evaporation of
cadmium and migration of tellurium to the damaged sites takes
place due to the laser-surface
interaction.~\cite{Hawkins_2008_JEMS,Teague_2009_JEMS}
%\section{Discussion} \label{sec:2}
The results on the photon induced nucleation of selenium
vacancies in TiSe$_2$ have broader implications for both
fundamental studies of TiSe$_2$ as well as potential practical
applications. The interest in TiSe$_2$ dichalcogenide has been due
to its excitonic nature of charge density wave state,
superconductivity in Cu$_x$TiSe$_2$~\cite{Morosan_2006_NaturePhys} and
Pd$_x$TiSe$_2$,~\cite{Morosan_2010_PRB} as well as due to unique
observation of the chiral charge density waves. Fundamental
studies of these correlated states and their coexistence have been
on the rise recently, as the dichalcogenide system is structurally
relatively simple and can provide some answers in regards to the
mechanisms of correlated electron states in other classes of
materials like perovskite high temperature superconductors or
pnictides. Since TiSe$_2$ is a compensated semimetal with very
large Hall coefficient at low temperatures~\cite{Salvo_1976_PRB},
the correlated electron states are very susceptible to the level
of intrinsic doping. Therefore, any fundamental studies should
take into account the selenium vacancy concentration levels
and the relatively low activation energy for their
nucleation~\cite{Novello_2015_PRB}. The concentration and dynamics of
selenium vacancies could be very important in TiSe$_2$ thin
films~\cite{Goli_2012_NanoLet,Li_2016_Nature,Peng_2015_PRB}
as the Raman peaks associated with these defects are strong in
synthesized films~\cite{Goli_2012_NanoLet,Bhatt_2014_ACSAMI}. With reduced
dimensionality of the system (from 3D crystals to 2D thin films),
the activation energy for selenium vacancy formation could further
decrease and make the 2D system metastable even at room
temperatures. The intrinsic doping due to vacancies might suppress
the charge density wave correlations and reduce the signature of
charge density wave phase normally observed resistivity vs
temperature measurements~\cite{Goli_2012_NanoLet}. A possible method to prevent selenium vaporization and at the same time enhance the CDW transition temperature in 2D TiSe$_2$ could be the encapsulation using hexagonal boron nitride, as shown recently.~\cite{Li_2016_APL} Clearly, more
work needs to be done in this area to understand the dynamics of
selenium in TiSe$_2$ and with recent advances in tip-enhanced Raman spectroscopy~\cite{Jiang_2016_ChemPhysLett} one could possibly obtain sufficient spatial and spectroscopic resolution.

%\section{Conclusions}
Raman spectroscopy studies on single crystal
TiSe$_2$ show a photon-induced defect signature associated with
nucleation of selenium vacancies. Two additional E$_g$ phonon
peaks at 241 and 263\,cm$^{-1}$ are observed in the laser
irradiated regions, consistent with Raman spectra of selenium
deficient TiSe$_2$. Temperature dependent studies of the threshold
laser intensity necessary to form selenium vacancies show that
there is a linear dependence, suggesting thermally activated
character of the process. The extrapolation of the onset
temperature of the photon-induced damage shows that the process of
selenium vacancy nucleation could start at temperatures as low as
435 K without the assistance of any photon sources.

\begin{acknowledgements} We would like to acknowledge the support
    by National Science Foundation under Grant No. ECCS-1408151.
    The use of the Center for Nanoscale
    Materials, an Office of Science user facility, was supported by
    the U. S. Department of Energy, Office of Science, Office of Basic
    Energy Sciences, under Contract No. DE-AC02-06CH11357.
\end{acknowledgements}

%\bibliographystyle{aipnum4-1}
%\bibliography{Lioi_AIP_v3}

\begin{thebibliography}{40}%
\makeatletter
\providecommand \@ifxundefined [1]{%
 \@ifx{#1\undefined}
}%
\providecommand \@ifnum [1]{%
 \ifnum #1\expandafter \@firstoftwo
 \else \expandafter \@secondoftwo
 \fi
}%
\providecommand \@ifx [1]{%
 \ifx #1\expandafter \@firstoftwo
 \else \expandafter \@secondoftwo
 \fi
}%
\providecommand \natexlab [1]{#1}%
\providecommand \enquote  [1]{``#1''}%
\providecommand \bibnamefont  [1]{#1}%
\providecommand \bibfnamefont [1]{#1}%
\providecommand \citenamefont [1]{#1}%
\providecommand \href@noop [0]{\@secondoftwo}%
\providecommand \href [0]{\begingroup \@sanitize@url \@href}%
\providecommand \@href[1]{\@@startlink{#1}\@@href}%
\providecommand \@@href[1]{\endgroup#1\@@endlink}%
\providecommand \@sanitize@url [0]{\catcode `\\12\catcode `\$12\catcode
  `\&12\catcode `\#12\catcode `\^12\catcode `\_12\catcode `\%12\relax}%
\providecommand \@@startlink[1]{}%
\providecommand \@@endlink[0]{}%
\providecommand \url  [0]{\begingroup\@sanitize@url \@url }%
\providecommand \@url [1]{\endgroup\@href {#1}{\urlprefix }}%
\providecommand \urlprefix  [0]{URL }%
\providecommand \Eprint [0]{\href }%
\providecommand \doibase [0]{http://dx.doi.org/}%
\providecommand \selectlanguage [0]{\@gobble}%
\providecommand \bibinfo  [0]{\@secondoftwo}%
\providecommand \bibfield  [0]{\@secondoftwo}%
\providecommand \translation [1]{[#1]}%
\providecommand \BibitemOpen [0]{}%
\providecommand \bibitemStop [0]{}%
\providecommand \bibitemNoStop [0]{.\EOS\space}%
\providecommand \EOS [0]{\spacefactor3000\relax}%
\providecommand \BibitemShut  [1]{\csname bibitem#1\endcsname}%
\let\auto@bib@innerbib\@empty
%</preamble>
\bibitem [{\citenamefont {Di~Salvo}, \citenamefont {Moncton},\ and\
  \citenamefont {Waszczak}(1976)}]{Salvo_1976_PRB}%
  \BibitemOpen
  \bibfield  {author} {\bibinfo {author} {\bibfnamefont {F.~J.}\ \bibnamefont
  {Di~Salvo}}, \bibinfo {author} {\bibfnamefont {D.~E.}\ \bibnamefont
  {Moncton}}, \ and\ \bibinfo {author} {\bibfnamefont {J.~V.}\ \bibnamefont
  {Waszczak}},\ }\href {http://link.aps.org/doi/10.1103/PhysRevB.14.4321}
  {\bibfield  {journal} {\bibinfo  {journal} {Phys. Rev. B}\ }\textbf {\bibinfo
  {volume} {14}},\ \bibinfo {pages} {4321} (\bibinfo {year}
  {1976})}\BibitemShut {NoStop}%
\bibitem [{\citenamefont {Wilson}(1978)}]{Wilson_1978_PSSB}%
  \BibitemOpen
  \bibfield  {author} {\bibinfo {author} {\bibfnamefont {J.~A.}\ \bibnamefont
  {Wilson}},\ }\href {\doibase 10.1002/pssb.2220860102} {\bibfield  {journal}
  {\bibinfo  {journal} {Phys. Status Solidi B}\ }\textbf {\bibinfo {volume}
  {86}},\ \bibinfo {pages} {11} (\bibinfo {year} {1978})}\BibitemShut {NoStop}%
\bibitem [{\citenamefont {Guo}\ \emph {et~al.}(2015)\citenamefont {Guo},
  \citenamefont {Yang}, \citenamefont {Yamamoto}, \citenamefont {Zhou},
  \citenamefont {Ishikawa}, \citenamefont {Ueno}, \citenamefont {Tsukagoshi},
  \citenamefont {Zhang}, \citenamefont {Dresselhaus},\ and\ \citenamefont
  {Saito}}]{Guo_2015_PRB}%
  \BibitemOpen
  \bibfield  {author} {\bibinfo {author} {\bibfnamefont {H.}~\bibnamefont
  {Guo}}, \bibinfo {author} {\bibfnamefont {T.}~\bibnamefont {Yang}}, \bibinfo
  {author} {\bibfnamefont {M.}~\bibnamefont {Yamamoto}}, \bibinfo {author}
  {\bibfnamefont {L.}~\bibnamefont {Zhou}}, \bibinfo {author} {\bibfnamefont
  {R.}~\bibnamefont {Ishikawa}}, \bibinfo {author} {\bibfnamefont
  {K.}~\bibnamefont {Ueno}}, \bibinfo {author} {\bibfnamefont {K.}~\bibnamefont
  {Tsukagoshi}}, \bibinfo {author} {\bibfnamefont {Z.}~\bibnamefont {Zhang}},
  \bibinfo {author} {\bibfnamefont {M.~S.}\ \bibnamefont {Dresselhaus}}, \ and\
  \bibinfo {author} {\bibfnamefont {R.}~\bibnamefont {Saito}},\ }\href
  {\doibase 10.1103/PhysRevB.91.205415} {\bibfield  {journal} {\bibinfo
  {journal} {Phys. Rev. B}\ }\textbf {\bibinfo {volume} {91}},\ \bibinfo
  {pages} {205415} (\bibinfo {year} {2015})}\BibitemShut {NoStop}%
\bibitem [{\citenamefont {Lorchat}, \citenamefont {Froehlicher},\ and\
  \citenamefont {Berciaud}(2016)}]{Lorchat_2016_ACSNano}%
  \BibitemOpen
  \bibfield  {author} {\bibinfo {author} {\bibfnamefont {E.}~\bibnamefont
  {Lorchat}}, \bibinfo {author} {\bibfnamefont {G.}~\bibnamefont
  {Froehlicher}}, \ and\ \bibinfo {author} {\bibfnamefont {S.}~\bibnamefont
  {Berciaud}},\ }\href {\doibase 10.1021/acsnano.5b07844} {\bibfield  {journal}
  {\bibinfo  {journal} {ACS Nano}\ }\textbf {\bibinfo {volume} {10}},\ \bibinfo
  {pages} {2752} (\bibinfo {year} {2016})}\BibitemShut {NoStop}%
\bibitem [{\citenamefont {Cercellier}\ \emph {et~al.}(2007)\citenamefont
  {Cercellier}, \citenamefont {Monney}, \citenamefont {Clerc}, \citenamefont
  {Battaglia}, \citenamefont {Despont}, \citenamefont {Garnier}, \citenamefont
  {Beck}, \citenamefont {Aebi}, \citenamefont {Patthey}, \citenamefont
  {Berger},\ and\ \citenamefont {Forro}}]{Cercellier_2007_PRL}%
  \BibitemOpen
  \bibfield  {author} {\bibinfo {author} {\bibfnamefont {H.}~\bibnamefont
  {Cercellier}}, \bibinfo {author} {\bibfnamefont {C.}~\bibnamefont {Monney}},
  \bibinfo {author} {\bibfnamefont {F.}~\bibnamefont {Clerc}}, \bibinfo
  {author} {\bibfnamefont {C.}~\bibnamefont {Battaglia}}, \bibinfo {author}
  {\bibfnamefont {L.}~\bibnamefont {Despont}}, \bibinfo {author} {\bibfnamefont
  {M.~G.}\ \bibnamefont {Garnier}}, \bibinfo {author} {\bibfnamefont
  {H.}~\bibnamefont {Beck}}, \bibinfo {author} {\bibfnamefont {P.}~\bibnamefont
  {Aebi}}, \bibinfo {author} {\bibfnamefont {L.}~\bibnamefont {Patthey}},
  \bibinfo {author} {\bibfnamefont {H.}~\bibnamefont {Berger}}, \ and\ \bibinfo
  {author} {\bibfnamefont {L.}~\bibnamefont {Forro}},\ }\href {\doibase
  10.1103/PhysRevLett.99.146403} {\bibfield  {journal} {\bibinfo  {journal}
  {Phys. Rev. Lett.}\ }\textbf {\bibinfo {volume} {99}},\ \bibinfo {pages}
  {146403} (\bibinfo {year} {2007})}\BibitemShut {NoStop}%
\bibitem [{\citenamefont {Monney}\ \emph {et~al.}(2010)\citenamefont {Monney},
  \citenamefont {Schwier}, \citenamefont {Garnier}, \citenamefont {Mariotti},
  \citenamefont {Didiot}, \citenamefont {Cercellier}, \citenamefont {Marcus},
  \citenamefont {Berger}, \citenamefont {Titov}, \citenamefont {Beck},\ and\
  \citenamefont {Aebi}}]{Monney_2010_NJP}%
  \BibitemOpen
  \bibfield  {author} {\bibinfo {author} {\bibfnamefont {C.}~\bibnamefont
  {Monney}}, \bibinfo {author} {\bibfnamefont {E.~F.}\ \bibnamefont {Schwier}},
  \bibinfo {author} {\bibfnamefont {M.~G.}\ \bibnamefont {Garnier}}, \bibinfo
  {author} {\bibfnamefont {N.}~\bibnamefont {Mariotti}}, \bibinfo {author}
  {\bibfnamefont {C.}~\bibnamefont {Didiot}}, \bibinfo {author} {\bibfnamefont
  {H.}~\bibnamefont {Cercellier}}, \bibinfo {author} {\bibfnamefont
  {J.}~\bibnamefont {Marcus}}, \bibinfo {author} {\bibfnamefont
  {H.}~\bibnamefont {Berger}}, \bibinfo {author} {\bibfnamefont {A.~N.}\
  \bibnamefont {Titov}}, \bibinfo {author} {\bibfnamefont {H.}~\bibnamefont
  {Beck}}, \ and\ \bibinfo {author} {\bibfnamefont {P.}~\bibnamefont {Aebi}},\
  }\href {\doibase 10.1088/1367-2630/12/12/125019} {\bibfield  {journal}
  {\bibinfo  {journal} {New J. Phys.}\ }\textbf {\bibinfo {volume} {12}},\
  \bibinfo {pages} {125019} (\bibinfo {year} {2010})}\BibitemShut {NoStop}%
\bibitem [{\citenamefont {Fogler}, \citenamefont {Butov},\ and\ \citenamefont
  {Novoselov}(2014)}]{Fogler_2014_NatureCom}%
  \BibitemOpen
  \bibfield  {author} {\bibinfo {author} {\bibfnamefont {M.~M.}\ \bibnamefont
  {Fogler}}, \bibinfo {author} {\bibfnamefont {L.~V.}\ \bibnamefont {Butov}}, \
  and\ \bibinfo {author} {\bibfnamefont {K.~S.}\ \bibnamefont {Novoselov}},\
  }\href {\doibase 10.1038/ncomms5555} {\bibfield  {journal} {\bibinfo
  {journal} {Nature Commun.}\ }\textbf {\bibinfo {volume} {5}},\ \bibinfo
  {pages} {4555} (\bibinfo {year} {2014})}\BibitemShut {NoStop}%
\bibitem [{\citenamefont {Korn}\ \emph {et~al.}(2011)\citenamefont {Korn},
  \citenamefont {Heydrich}, \citenamefont {Hirmer}, \citenamefont
  {Schmutzler},\ and\ \citenamefont {Schueller}}]{Korn_2011_APL}%
  \BibitemOpen
  \bibfield  {author} {\bibinfo {author} {\bibfnamefont {T.}~\bibnamefont
  {Korn}}, \bibinfo {author} {\bibfnamefont {S.}~\bibnamefont {Heydrich}},
  \bibinfo {author} {\bibfnamefont {M.}~\bibnamefont {Hirmer}}, \bibinfo
  {author} {\bibfnamefont {J.}~\bibnamefont {Schmutzler}}, \ and\ \bibinfo
  {author} {\bibfnamefont {C.}~\bibnamefont {Schueller}},\ }\href {\doibase
  10.1063/1.3636402} {\bibfield  {journal} {\bibinfo  {journal} {Appl. Phys.
  Lett.}\ }\textbf {\bibinfo {volume} {99}},\ \bibinfo {pages} {102109}
  (\bibinfo {year} {2011})}\BibitemShut {NoStop}%
\bibitem [{\citenamefont {Bhatt}\ \emph {et~al.}(2013)\citenamefont {Bhatt},
  \citenamefont {Basu}, \citenamefont {Bhattacharya}, \citenamefont {Singh},
  \citenamefont {Aswal}, \citenamefont {Gupta}, \citenamefont {Okram},
  \citenamefont {Ganesan}, \citenamefont {Venkateshwarlu}, \citenamefont
  {Surgers}, \citenamefont {Navaneethan},\ and\ \citenamefont
  {Hayakawa}}]{Bhatt_2013_APMSP}%
  \BibitemOpen
  \bibfield  {author} {\bibinfo {author} {\bibfnamefont {R.}~\bibnamefont
  {Bhatt}}, \bibinfo {author} {\bibfnamefont {R.}~\bibnamefont {Basu}},
  \bibinfo {author} {\bibfnamefont {S.}~\bibnamefont {Bhattacharya}}, \bibinfo
  {author} {\bibfnamefont {A.}~\bibnamefont {Singh}}, \bibinfo {author}
  {\bibfnamefont {D.~K.}\ \bibnamefont {Aswal}}, \bibinfo {author}
  {\bibfnamefont {S.~K.}\ \bibnamefont {Gupta}}, \bibinfo {author}
  {\bibfnamefont {G.~S.}\ \bibnamefont {Okram}}, \bibinfo {author}
  {\bibfnamefont {V.}~\bibnamefont {Ganesan}}, \bibinfo {author} {\bibfnamefont
  {D.}~\bibnamefont {Venkateshwarlu}}, \bibinfo {author} {\bibfnamefont
  {C.}~\bibnamefont {Surgers}}, \bibinfo {author} {\bibfnamefont
  {M.}~\bibnamefont {Navaneethan}}, \ and\ \bibinfo {author} {\bibfnamefont
  {Y.}~\bibnamefont {Hayakawa}},\ }\href {\doibase 10.1007/s00339-012-7536-8}
  {\bibfield  {journal} {\bibinfo  {journal} {Appl. Phys. A-Materials Science
  and Processing}\ }\textbf {\bibinfo {volume} {111}},\ \bibinfo {pages} {465}
  (\bibinfo {year} {2013})}\BibitemShut {NoStop}%
\bibitem [{\citenamefont {Ishioka}\ \emph {et~al.}(2010)\citenamefont
  {Ishioka}, \citenamefont {Liu}, \citenamefont {Shimatake}, \citenamefont
  {Kurosawa}, \citenamefont {Ichimura}, \citenamefont {Toda}, \citenamefont
  {Oda},\ and\ \citenamefont {Tanda}}]{Ishioka_2010_PRL}%
  \BibitemOpen
  \bibfield  {author} {\bibinfo {author} {\bibfnamefont {J.}~\bibnamefont
  {Ishioka}}, \bibinfo {author} {\bibfnamefont {Y.~H.}\ \bibnamefont {Liu}},
  \bibinfo {author} {\bibfnamefont {K.}~\bibnamefont {Shimatake}}, \bibinfo
  {author} {\bibfnamefont {T.}~\bibnamefont {Kurosawa}}, \bibinfo {author}
  {\bibfnamefont {K.}~\bibnamefont {Ichimura}}, \bibinfo {author}
  {\bibfnamefont {Y.}~\bibnamefont {Toda}}, \bibinfo {author} {\bibfnamefont
  {M.}~\bibnamefont {Oda}}, \ and\ \bibinfo {author} {\bibfnamefont
  {S.}~\bibnamefont {Tanda}},\ }\href
  {http://link.aps.org/doi/10.1103/PhysRevLett.105.176401} {\bibfield
  {journal} {\bibinfo  {journal} {Phys. Rev. Lett.}\ }\textbf {\bibinfo
  {volume} {105}},\ \bibinfo {pages} {176401} (\bibinfo {year}
  {2010})}\BibitemShut {NoStop}%
\bibitem [{\citenamefont {Castellan}\ \emph {et~al.}(2013)\citenamefont
  {Castellan}, \citenamefont {Rosenkranz}, \citenamefont {Osborn},
  \citenamefont {Li}, \citenamefont {Gray}, \citenamefont {Luo}, \citenamefont
  {Welp}, \citenamefont {Karapetrov}, \citenamefont {Ruff},\ and\ \citenamefont
  {van Wezel}}]{Castellan_2013_PRL}%
  \BibitemOpen
  \bibfield  {author} {\bibinfo {author} {\bibfnamefont {J.-P.}\ \bibnamefont
  {Castellan}}, \bibinfo {author} {\bibfnamefont {S.}~\bibnamefont
  {Rosenkranz}}, \bibinfo {author} {\bibfnamefont {R.}~\bibnamefont {Osborn}},
  \bibinfo {author} {\bibfnamefont {Q.}~\bibnamefont {Li}}, \bibinfo {author}
  {\bibfnamefont {K.~E.}\ \bibnamefont {Gray}}, \bibinfo {author}
  {\bibfnamefont {X.}~\bibnamefont {Luo}}, \bibinfo {author} {\bibfnamefont
  {U.}~\bibnamefont {Welp}}, \bibinfo {author} {\bibfnamefont {G.}~\bibnamefont
  {Karapetrov}}, \bibinfo {author} {\bibfnamefont {J.~P.~C.}\ \bibnamefont
  {Ruff}}, \ and\ \bibinfo {author} {\bibfnamefont {J.}~\bibnamefont {van
  Wezel}},\ }\href {\doibase 10.1103/PhysRevLett.110.196404} {\bibfield
  {journal} {\bibinfo  {journal} {Phys. Rev. Lett.}\ }\textbf {\bibinfo
  {volume} {110}},\ \bibinfo {pages} {196404} (\bibinfo {year}
  {2013})}\BibitemShut {NoStop}%
\bibitem [{\citenamefont {Chen}\ \emph {et~al.}(2015)\citenamefont {Chen},
  \citenamefont {Chan}, \citenamefont {Fang}, \citenamefont {Zhang},
  \citenamefont {Chou}, \citenamefont {Mo}, \citenamefont {Hussain},
  \citenamefont {Fedorov},\ and\ \citenamefont
  {Chiang}}]{TCChiang_2015_NatComm}%
  \BibitemOpen
  \bibfield  {author} {\bibinfo {author} {\bibfnamefont {P.}~\bibnamefont
  {Chen}}, \bibinfo {author} {\bibfnamefont {Y.~H.}\ \bibnamefont {Chan}},
  \bibinfo {author} {\bibfnamefont {X.~Y.}\ \bibnamefont {Fang}}, \bibinfo
  {author} {\bibfnamefont {Y.}~\bibnamefont {Zhang}}, \bibinfo {author}
  {\bibfnamefont {M.~Y.}\ \bibnamefont {Chou}}, \bibinfo {author}
  {\bibfnamefont {S.~K.}\ \bibnamefont {Mo}}, \bibinfo {author} {\bibfnamefont
  {Z.}~\bibnamefont {Hussain}}, \bibinfo {author} {\bibfnamefont {A.~V.}\
  \bibnamefont {Fedorov}}, \ and\ \bibinfo {author} {\bibfnamefont {T.~C.}\
  \bibnamefont {Chiang}},\ }\href {\doibase 10.1038/ncomms9943} {\bibfield
  {journal} {\bibinfo  {journal} {Nature Commun.}\ }\textbf {\bibinfo {volume}
  {6}},\ \bibinfo {pages} {8943} (\bibinfo {year} {2015})}\BibitemShut
  {NoStop}%
\bibitem [{\citenamefont {Li}\ \emph {et~al.}(2016{\natexlab{a}})\citenamefont
  {Li}, \citenamefont {O'Farrell}, \citenamefont {Loh}, \citenamefont {Eda},
  \citenamefont {Ozyilmaz},\ and\ \citenamefont {Neto}}]{Li_2016_Nature}%
  \BibitemOpen
  \bibfield  {author} {\bibinfo {author} {\bibfnamefont {L.~J.}\ \bibnamefont
  {Li}}, \bibinfo {author} {\bibfnamefont {E.~C.~T.}\ \bibnamefont
  {O'Farrell}}, \bibinfo {author} {\bibfnamefont {K.~P.}\ \bibnamefont {Loh}},
  \bibinfo {author} {\bibfnamefont {G.}~\bibnamefont {Eda}}, \bibinfo {author}
  {\bibfnamefont {B.}~\bibnamefont {Ozyilmaz}}, \ and\ \bibinfo {author}
  {\bibfnamefont {A.~H.~C.}\ \bibnamefont {Neto}},\ }\href {\doibase
  10.1038/nature16175} {\bibfield  {journal} {\bibinfo  {journal} {Nature}\
  }\textbf {\bibinfo {volume} {529}},\ \bibinfo {pages} {185} (\bibinfo {year}
  {2016}{\natexlab{a}})}\BibitemShut {NoStop}%
\bibitem [{\citenamefont {Bhatt}\ \emph {et~al.}(2014)\citenamefont {Bhatt},
  \citenamefont {Bhattacharya}, \citenamefont {Basu}, \citenamefont {Ahmad},
  \citenamefont {Chauhan}, \citenamefont {Okram}, \citenamefont {Bhatt},
  \citenamefont {Roy}, \citenamefont {Navaneethan}, \citenamefont {Hayakawa},
  \citenamefont {Debnath}, \citenamefont {Singh}, \citenamefont {Aswal},\ and\
  \citenamefont {Gupta}}]{Bhatt_2014_ACSAMI}%
  \BibitemOpen
  \bibfield  {author} {\bibinfo {author} {\bibfnamefont {R.}~\bibnamefont
  {Bhatt}}, \bibinfo {author} {\bibfnamefont {S.}~\bibnamefont {Bhattacharya}},
  \bibinfo {author} {\bibfnamefont {R.}~\bibnamefont {Basu}}, \bibinfo {author}
  {\bibfnamefont {S.}~\bibnamefont {Ahmad}}, \bibinfo {author} {\bibfnamefont
  {A.~K.}\ \bibnamefont {Chauhan}}, \bibinfo {author} {\bibfnamefont {G.~S.}\
  \bibnamefont {Okram}}, \bibinfo {author} {\bibfnamefont {P.}~\bibnamefont
  {Bhatt}}, \bibinfo {author} {\bibfnamefont {M.}~\bibnamefont {Roy}}, \bibinfo
  {author} {\bibfnamefont {M.}~\bibnamefont {Navaneethan}}, \bibinfo {author}
  {\bibfnamefont {Y.}~\bibnamefont {Hayakawa}}, \bibinfo {author}
  {\bibfnamefont {A.~K.}\ \bibnamefont {Debnath}}, \bibinfo {author}
  {\bibfnamefont {A.}~\bibnamefont {Singh}}, \bibinfo {author} {\bibfnamefont
  {D.~K.}\ \bibnamefont {Aswal}}, \ and\ \bibinfo {author} {\bibfnamefont
  {S.~K.}\ \bibnamefont {Gupta}},\ }\href {\doibase 10.1021/am503477z}
  {\bibfield  {journal} {\bibinfo  {journal} {ACS Appl. Mater. Interfaces}\
  }\textbf {\bibinfo {volume} {6}},\ \bibinfo {pages} {18619} (\bibinfo {year}
  {2014})}\BibitemShut {NoStop}%
\bibitem [{\citenamefont {Gu}\ \emph {et~al.}(2015)\citenamefont {Gu},
  \citenamefont {Katsura}, \citenamefont {Yoshino}, \citenamefont {Takagi},\
  and\ \citenamefont {Taniguchi}}]{Gu_2015_ScientificRep}%
  \BibitemOpen
  \bibfield  {author} {\bibinfo {author} {\bibfnamefont {Y.}~\bibnamefont
  {Gu}}, \bibinfo {author} {\bibfnamefont {Y.}~\bibnamefont {Katsura}},
  \bibinfo {author} {\bibfnamefont {T.}~\bibnamefont {Yoshino}}, \bibinfo
  {author} {\bibfnamefont {H.}~\bibnamefont {Takagi}}, \ and\ \bibinfo {author}
  {\bibfnamefont {K.}~\bibnamefont {Taniguchi}},\ }\href {\doibase
  10.1038/srep12486} {\bibfield  {journal} {\bibinfo  {journal} {Sci. Rep.}\
  }\textbf {\bibinfo {volume} {5}},\ \bibinfo {pages} {12486} (\bibinfo {year}
  {2015})}\BibitemShut {NoStop}%
\bibitem [{\citenamefont {Lv}\ \emph {et~al.}(2015)\citenamefont {Lv},
  \citenamefont {Robinson}, \citenamefont {Schaak}, \citenamefont {Sun},
  \citenamefont {Sun}, \citenamefont {Mallouk},\ and\ \citenamefont
  {Terrones}}]{Lv_2015_ACR}%
  \BibitemOpen
  \bibfield  {author} {\bibinfo {author} {\bibfnamefont {R.}~\bibnamefont
  {Lv}}, \bibinfo {author} {\bibfnamefont {J.~A.}\ \bibnamefont {Robinson}},
  \bibinfo {author} {\bibfnamefont {R.~E.}\ \bibnamefont {Schaak}}, \bibinfo
  {author} {\bibfnamefont {D.}~\bibnamefont {Sun}}, \bibinfo {author}
  {\bibfnamefont {Y.}~\bibnamefont {Sun}}, \bibinfo {author} {\bibfnamefont
  {T.~E.}\ \bibnamefont {Mallouk}}, \ and\ \bibinfo {author} {\bibfnamefont
  {M.}~\bibnamefont {Terrones}},\ }\href {\doibase 10.1021/ar5002846}
  {\bibfield  {journal} {\bibinfo  {journal} {Acc. Chem. Res.}\ }\textbf
  {\bibinfo {volume} {48}},\ \bibinfo {pages} {56} (\bibinfo {year}
  {2015})}\BibitemShut {NoStop}%
\bibitem [{\citenamefont {Zhu}, \citenamefont {Cheng},\ and\ \citenamefont
  {Schwingenschlogl}(2011)}]{Zhu_2011_PRB}%
  \BibitemOpen
  \bibfield  {author} {\bibinfo {author} {\bibfnamefont {Z.~Y.}\ \bibnamefont
  {Zhu}}, \bibinfo {author} {\bibfnamefont {Y.~C.}\ \bibnamefont {Cheng}}, \
  and\ \bibinfo {author} {\bibfnamefont {U.}~\bibnamefont {Schwingenschlogl}},\
  }\href {\doibase 10.1103/PhysRevB.84.153402} {\bibfield  {journal} {\bibinfo
  {journal} {Phys. Rev. B}\ }\textbf {\bibinfo {volume} {84}},\ \bibinfo
  {pages} {153402} (\bibinfo {year} {2011})}\BibitemShut {NoStop}%
\bibitem [{\citenamefont {Lebegue}\ and\ \citenamefont
  {Eriksson}(2009)}]{Lebegue_2009_PRB}%
  \BibitemOpen
  \bibfield  {author} {\bibinfo {author} {\bibfnamefont {S.}~\bibnamefont
  {Lebegue}}\ and\ \bibinfo {author} {\bibfnamefont {O.}~\bibnamefont
  {Eriksson}},\ }\href {\doibase 10.1103/PhysRevB.79.115409} {\bibfield
  {journal} {\bibinfo  {journal} {Phys. Rev. B}\ }\textbf {\bibinfo {volume}
  {79}},\ \bibinfo {pages} {115409} (\bibinfo {year} {2009})}\BibitemShut
  {NoStop}%
\bibitem [{\citenamefont {Ubaldini}\ and\ \citenamefont
  {Giannini}(2014)}]{Ulbaldini_2014_JoCG}%
  \BibitemOpen
  \bibfield  {author} {\bibinfo {author} {\bibfnamefont {A.}~\bibnamefont
  {Ubaldini}}\ and\ \bibinfo {author} {\bibfnamefont {E.}~\bibnamefont
  {Giannini}},\ }\href {\doibase 10.1016/j.jcrysgro.2014.12.070} {\bibfield
  {journal} {\bibinfo  {journal} {J. Cryst. Growth}\ }\textbf {\bibinfo
  {volume} {401}},\ \bibinfo {pages} {878} (\bibinfo {year}
  {2014})}\BibitemShut {NoStop}%
\bibitem [{\citenamefont {Peng}\ \emph {et~al.}(2015)\citenamefont {Peng},
  \citenamefont {Guan}, \citenamefont {Zhang}, \citenamefont {Song},
  \citenamefont {Wang}, \citenamefont {He}, \citenamefont {Xue},\ and\
  \citenamefont {Ma}}]{Peng_2015_PRB}%
  \BibitemOpen
  \bibfield  {author} {\bibinfo {author} {\bibfnamefont {J.-P.}\ \bibnamefont
  {Peng}}, \bibinfo {author} {\bibfnamefont {J.-Q.}\ \bibnamefont {Guan}},
  \bibinfo {author} {\bibfnamefont {H.-M.}\ \bibnamefont {Zhang}}, \bibinfo
  {author} {\bibfnamefont {C.-L.}\ \bibnamefont {Song}}, \bibinfo {author}
  {\bibfnamefont {L.}~\bibnamefont {Wang}}, \bibinfo {author} {\bibfnamefont
  {K.}~\bibnamefont {He}}, \bibinfo {author} {\bibfnamefont {Q.-K.}\
  \bibnamefont {Xue}}, \ and\ \bibinfo {author} {\bibfnamefont {X.-C.}\
  \bibnamefont {Ma}},\ }\href {\doibase 10.1103/PhysRevB.91.121113} {\bibfield
  {journal} {\bibinfo  {journal} {Phys. Rev. B}\ }\textbf {\bibinfo {volume}
  {91}},\ \bibinfo {pages} {121113(R)} (\bibinfo {year} {2015})}\BibitemShut
  {NoStop}%
\bibitem [{\citenamefont {Boscher}, \citenamefont {Carmalt},\ and\
  \citenamefont {Parkin}(2006)}]{Boscher_2006_CVD}%
  \BibitemOpen
  \bibfield  {author} {\bibinfo {author} {\bibfnamefont {N.~D.}\ \bibnamefont
  {Boscher}}, \bibinfo {author} {\bibfnamefont {C.~J.}\ \bibnamefont
  {Carmalt}}, \ and\ \bibinfo {author} {\bibfnamefont {I.~P.}\ \bibnamefont
  {Parkin}},\ }\href {\doibase 10.1002/cvde.200506423} {\bibfield  {journal}
  {\bibinfo  {journal} {Chem. Vap. Deposition}\ }\textbf {\bibinfo {volume}
  {12}},\ \bibinfo {pages} {54} (\bibinfo {year} {2006})}\BibitemShut {NoStop}%
\bibitem [{\citenamefont {Sugawara}\ \emph {et~al.}(2016)\citenamefont
  {Sugawara}, \citenamefont {Nakata}, \citenamefont {Shimizu}, \citenamefont
  {Han}, \citenamefont {Hitosugi}, \citenamefont {Sato},\ and\ \citenamefont
  {Takahashi}}]{Sugawara_2016_ACSNano}%
  \BibitemOpen
  \bibfield  {author} {\bibinfo {author} {\bibfnamefont {K.}~\bibnamefont
  {Sugawara}}, \bibinfo {author} {\bibfnamefont {Y.}~\bibnamefont {Nakata}},
  \bibinfo {author} {\bibfnamefont {R.}~\bibnamefont {Shimizu}}, \bibinfo
  {author} {\bibfnamefont {P.}~\bibnamefont {Han}}, \bibinfo {author}
  {\bibfnamefont {T.}~\bibnamefont {Hitosugi}}, \bibinfo {author}
  {\bibfnamefont {T.}~\bibnamefont {Sato}}, \ and\ \bibinfo {author}
  {\bibfnamefont {T.}~\bibnamefont {Takahashi}},\ }\href {\doibase
  10.1021/acsnano.5b06727} {\bibfield  {journal} {\bibinfo  {journal} {ACS
  Nano}\ }\textbf {\bibinfo {volume} {10}},\ \bibinfo {pages} {1341} (\bibinfo
  {year} {2016})}\BibitemShut {NoStop}%
\bibitem [{\citenamefont {Hildebrand}\ \emph {et~al.}(2014)\citenamefont
  {Hildebrand}, \citenamefont {Didiot}, \citenamefont {Novello}, \citenamefont
  {Monney}, \citenamefont {Scarfato}, \citenamefont {Ubaldini}, \citenamefont
  {Berger}, \citenamefont {Bowler}, \citenamefont {Renner},\ and\ \citenamefont
  {Aebi}}]{Hildebrand_2014_PRL}%
  \BibitemOpen
  \bibfield  {author} {\bibinfo {author} {\bibfnamefont {B.}~\bibnamefont
  {Hildebrand}}, \bibinfo {author} {\bibfnamefont {C.}~\bibnamefont {Didiot}},
  \bibinfo {author} {\bibfnamefont {A.~M.}\ \bibnamefont {Novello}}, \bibinfo
  {author} {\bibfnamefont {G.}~\bibnamefont {Monney}}, \bibinfo {author}
  {\bibfnamefont {A.}~\bibnamefont {Scarfato}}, \bibinfo {author}
  {\bibfnamefont {A.}~\bibnamefont {Ubaldini}}, \bibinfo {author}
  {\bibfnamefont {H.}~\bibnamefont {Berger}}, \bibinfo {author} {\bibfnamefont
  {D.~R.}\ \bibnamefont {Bowler}}, \bibinfo {author} {\bibfnamefont
  {C.}~\bibnamefont {Renner}}, \ and\ \bibinfo {author} {\bibfnamefont
  {P.}~\bibnamefont {Aebi}},\ }\href {\doibase 10.1103/PhysRevLett.112.197001}
  {\bibfield  {journal} {\bibinfo  {journal} {Phys. Rev. Lett.}\ }\textbf
  {\bibinfo {volume} {112}},\ \bibinfo {pages} {197001} (\bibinfo {year}
  {2014})}\BibitemShut {NoStop}%
\bibitem [{\citenamefont {Zhao}\ \emph {et~al.}(2011)\citenamefont {Zhao},
  \citenamefont {Tan}, \citenamefont {Liu},\ and\ \citenamefont
  {Ferrari}}]{Zhao_2011_JACS}%
  \BibitemOpen
  \bibfield  {author} {\bibinfo {author} {\bibfnamefont {W.}~\bibnamefont
  {Zhao}}, \bibinfo {author} {\bibfnamefont {P.~H.}\ \bibnamefont {Tan}},
  \bibinfo {author} {\bibfnamefont {J.}~\bibnamefont {Liu}}, \ and\ \bibinfo
  {author} {\bibfnamefont {A.~C.}\ \bibnamefont {Ferrari}},\ }\href {\doibase
  10.1021/ja110939a} {\bibfield  {journal} {\bibinfo  {journal} {J. Am. Chem.
  Soc.}\ }\textbf {\bibinfo {volume} {133}},\ \bibinfo {pages} {5941} (\bibinfo
  {year} {2011})}\BibitemShut {NoStop}%
\bibitem [{\citenamefont {Soubelet}\ \emph {et~al.}(2016)\citenamefont
  {Soubelet}, \citenamefont {Bruchhausen}, \citenamefont {Fainstein},
  \citenamefont {Nogajewski},\ and\ \citenamefont
  {Faugeras}}]{Soubelet_2016_PRB}%
  \BibitemOpen
  \bibfield  {author} {\bibinfo {author} {\bibfnamefont {P.}~\bibnamefont
  {Soubelet}}, \bibinfo {author} {\bibfnamefont {A.~E.}\ \bibnamefont
  {Bruchhausen}}, \bibinfo {author} {\bibfnamefont {A.}~\bibnamefont
  {Fainstein}}, \bibinfo {author} {\bibfnamefont {K.}~\bibnamefont
  {Nogajewski}}, \ and\ \bibinfo {author} {\bibfnamefont {C.}~\bibnamefont
  {Faugeras}},\ }\href {\doibase 10.1103/PhysRevB.93.155407} {\bibfield
  {journal} {\bibinfo  {journal} {Phys. Rev. B}\ }\textbf {\bibinfo {volume}
  {93}},\ \bibinfo {pages} {155407} (\bibinfo {year} {2016})}\BibitemShut
  {NoStop}%
\bibitem [{\citenamefont {Staiger}\ \emph {et~al.}(2015)\citenamefont
  {Staiger}, \citenamefont {Gillen}, \citenamefont {Scheuschner}, \citenamefont
  {Ochedowski}, \citenamefont {Kampmann}, \citenamefont {Schleberger},
  \citenamefont {Thomsen},\ and\ \citenamefont {Maultzsch}}]{Staiger_2015_PRB}%
  \BibitemOpen
  \bibfield  {author} {\bibinfo {author} {\bibfnamefont {M.}~\bibnamefont
  {Staiger}}, \bibinfo {author} {\bibfnamefont {R.}~\bibnamefont {Gillen}},
  \bibinfo {author} {\bibfnamefont {N.}~\bibnamefont {Scheuschner}}, \bibinfo
  {author} {\bibfnamefont {O.}~\bibnamefont {Ochedowski}}, \bibinfo {author}
  {\bibfnamefont {F.}~\bibnamefont {Kampmann}}, \bibinfo {author}
  {\bibfnamefont {M.}~\bibnamefont {Schleberger}}, \bibinfo {author}
  {\bibfnamefont {C.}~\bibnamefont {Thomsen}}, \ and\ \bibinfo {author}
  {\bibfnamefont {J.}~\bibnamefont {Maultzsch}},\ }\href {\doibase
  10.1103/PhysRevB.91.195419} {\bibfield  {journal} {\bibinfo  {journal} {Phys.
  Rev. B}\ }\textbf {\bibinfo {volume} {91}},\ \bibinfo {pages} {195419}
  (\bibinfo {year} {2015})}\BibitemShut {NoStop}%
\bibitem [{\citenamefont {Oglesby}\ \emph {et~al.}(1994)\citenamefont
  {Oglesby}, \citenamefont {Bucher}, \citenamefont {Kloc},\ and\ \citenamefont
  {Hohl}}]{Oglesby_1994_JoCG}%
  \BibitemOpen
  \bibfield  {author} {\bibinfo {author} {\bibfnamefont {C.~S.}\ \bibnamefont
  {Oglesby}}, \bibinfo {author} {\bibfnamefont {E.}~\bibnamefont {Bucher}},
  \bibinfo {author} {\bibfnamefont {C.}~\bibnamefont {Kloc}}, \ and\ \bibinfo
  {author} {\bibfnamefont {H.}~\bibnamefont {Hohl}},\ }\href {\doibase
  10.1016/0022-0248(94)91287-4} {\bibfield  {journal} {\bibinfo  {journal} {J.
  Cryst. Growth}\ }\textbf {\bibinfo {volume} {137}},\ \bibinfo {pages} {289}
  (\bibinfo {year} {1994})}\BibitemShut {NoStop}%
\bibitem [{\citenamefont {Iavarone}\ \emph {et~al.}(2012)\citenamefont
  {Iavarone}, \citenamefont {Di~Capua}, \citenamefont {Zhang}, \citenamefont
  {Golalikhani}, \citenamefont {Moore},\ and\ \citenamefont
  {Karapetrov}}]{Iavarone_2012_PRB}%
  \BibitemOpen
  \bibfield  {author} {\bibinfo {author} {\bibfnamefont {M.}~\bibnamefont
  {Iavarone}}, \bibinfo {author} {\bibfnamefont {R.}~\bibnamefont {Di~Capua}},
  \bibinfo {author} {\bibfnamefont {X.}~\bibnamefont {Zhang}}, \bibinfo
  {author} {\bibfnamefont {M.}~\bibnamefont {Golalikhani}}, \bibinfo {author}
  {\bibfnamefont {S.~A.}\ \bibnamefont {Moore}}, \ and\ \bibinfo {author}
  {\bibfnamefont {G.}~\bibnamefont {Karapetrov}},\ }\href {\doibase
  10.1103/PhysRevB.85.155103} {\bibfield  {journal} {\bibinfo  {journal} {Phys.
  Rev. B}\ }\textbf {\bibinfo {volume} {85}},\ \bibinfo {pages} {155103}
  (\bibinfo {year} {2012})}\BibitemShut {NoStop}%
\bibitem [{\citenamefont {Snow}\ \emph {et~al.}(2003)\citenamefont {Snow},
  \citenamefont {Karpus}, \citenamefont {Cooper}, \citenamefont {Kidd},\ and\
  \citenamefont {Chiang}}]{Snow_2003_PRL}%
  \BibitemOpen
  \bibfield  {author} {\bibinfo {author} {\bibfnamefont {C.~S.}\ \bibnamefont
  {Snow}}, \bibinfo {author} {\bibfnamefont {J.~F.}\ \bibnamefont {Karpus}},
  \bibinfo {author} {\bibfnamefont {S.~L.}\ \bibnamefont {Cooper}}, \bibinfo
  {author} {\bibfnamefont {T.~E.}\ \bibnamefont {Kidd}}, \ and\ \bibinfo
  {author} {\bibfnamefont {T.~C.}\ \bibnamefont {Chiang}},\ }\href {\doibase
  10.1103/PhysRevLett.91.136402} {\bibfield  {journal} {\bibinfo  {journal}
  {Phys. Rev. Lett.}\ }\textbf {\bibinfo {volume} {91}},\ \bibinfo {pages}
  {136402} (\bibinfo {year} {2003})}\BibitemShut {NoStop}%
\bibitem [{\citenamefont {Sugai}\ \emph {et~al.}(1980)\citenamefont {Sugai},
  \citenamefont {Murase}, \citenamefont {Uchida},\ and\ \citenamefont
  {Tanaka}}]{Sugai_1980_SSC}%
  \BibitemOpen
  \bibfield  {author} {\bibinfo {author} {\bibfnamefont {S.}~\bibnamefont
  {Sugai}}, \bibinfo {author} {\bibfnamefont {K.}~\bibnamefont {Murase}},
  \bibinfo {author} {\bibfnamefont {S.}~\bibnamefont {Uchida}}, \ and\ \bibinfo
  {author} {\bibfnamefont {S.}~\bibnamefont {Tanaka}},\ }\href {\doibase
  10.1016/0038-1098(80)90175-1} {\bibfield  {journal} {\bibinfo  {journal}
  {Solid State Commun.}\ }\textbf {\bibinfo {volume} {35}},\ \bibinfo {pages}
  {433} (\bibinfo {year} {1980})}\BibitemShut {NoStop}%
\bibitem [{\citenamefont {Holy}\ \emph {et~al.}(1977)\citenamefont {Holy},
  \citenamefont {Woo}, \citenamefont {Klein},\ and\ \citenamefont
  {Brown}}]{Holy_1977_PRB}%
  \BibitemOpen
  \bibfield  {author} {\bibinfo {author} {\bibfnamefont {J.~A.}\ \bibnamefont
  {Holy}}, \bibinfo {author} {\bibfnamefont {K.~C.}\ \bibnamefont {Woo}},
  \bibinfo {author} {\bibfnamefont {M.~V.}\ \bibnamefont {Klein}}, \ and\
  \bibinfo {author} {\bibfnamefont {F.~C.}\ \bibnamefont {Brown}},\ }\href
  {http://link.aps.org/doi/10.1103/PhysRevB.16.3628} {\bibfield  {journal}
  {\bibinfo  {journal} {Phys. Rev. B}\ }\textbf {\bibinfo {volume} {16}},\
  \bibinfo {pages} {3628} (\bibinfo {year} {1977})}\BibitemShut {NoStop}%
\bibitem [{\citenamefont {Barath}\ \emph {et~al.}(2008)\citenamefont {Barath},
  \citenamefont {Kim}, \citenamefont {Karpus}, \citenamefont {Cooper},
  \citenamefont {Abbamonte}, \citenamefont {Fradkin},\ and\ \citenamefont
  {Cava}}]{Barath_2008_PRL}%
  \BibitemOpen
  \bibfield  {author} {\bibinfo {author} {\bibfnamefont {H.}~\bibnamefont
  {Barath}}, \bibinfo {author} {\bibfnamefont {M.}~\bibnamefont {Kim}},
  \bibinfo {author} {\bibfnamefont {J.~F.}\ \bibnamefont {Karpus}}, \bibinfo
  {author} {\bibfnamefont {S.~L.}\ \bibnamefont {Cooper}}, \bibinfo {author}
  {\bibfnamefont {P.}~\bibnamefont {Abbamonte}}, \bibinfo {author}
  {\bibfnamefont {E.}~\bibnamefont {Fradkin}}, \ and\ \bibinfo {author}
  {\bibfnamefont {R.~J.}\ \bibnamefont {Cava}},\ }\href {\doibase
  10.1103/PhysRevLett.100.106402} {\bibfield  {journal} {\bibinfo  {journal}
  {Phys. Rev. Lett.}\ }\textbf {\bibinfo {volume} {100}},\ \bibinfo {pages}
  {106402} (\bibinfo {year} {2008})}\BibitemShut {NoStop}%
\bibitem [{\citenamefont {Hawkins}\ \emph {et~al.}(2008)\citenamefont
  {Hawkins}, \citenamefont {Villa-Aleman}, \citenamefont {Duff}, \citenamefont
  {Hunter}, \citenamefont {Burger}, \citenamefont {Groza}, \citenamefont
  {Buliga},\ and\ \citenamefont {Black}}]{Hawkins_2008_JEMS}%
  \BibitemOpen
  \bibfield  {author} {\bibinfo {author} {\bibfnamefont {S.~A.}\ \bibnamefont
  {Hawkins}}, \bibinfo {author} {\bibfnamefont {E.}~\bibnamefont
  {Villa-Aleman}}, \bibinfo {author} {\bibfnamefont {M.~C.}\ \bibnamefont
  {Duff}}, \bibinfo {author} {\bibfnamefont {D.~B.}\ \bibnamefont {Hunter}},
  \bibinfo {author} {\bibfnamefont {A.}~\bibnamefont {Burger}}, \bibinfo
  {author} {\bibfnamefont {M.}~\bibnamefont {Groza}}, \bibinfo {author}
  {\bibfnamefont {V.}~\bibnamefont {Buliga}}, \ and\ \bibinfo {author}
  {\bibfnamefont {D.~R.}\ \bibnamefont {Black}},\ }\href {\doibase
  10.1007/s11664-008-0448-x} {\bibfield  {journal} {\bibinfo  {journal} {J.
  Electron. Mater.}\ }\textbf {\bibinfo {volume} {37}},\ \bibinfo {pages}
  {1438} (\bibinfo {year} {2008})}\BibitemShut {NoStop}%
\bibitem [{\citenamefont {Teague}\ \emph {et~al.}(2009)\citenamefont {Teague},
  \citenamefont {Hawkins}, \citenamefont {Duff}, \citenamefont {Groza},
  \citenamefont {Buliga},\ and\ \citenamefont {Burger}}]{Teague_2009_JEMS}%
  \BibitemOpen
  \bibfield  {author} {\bibinfo {author} {\bibfnamefont {L.~C.}\ \bibnamefont
  {Teague}}, \bibinfo {author} {\bibfnamefont {S.~A.}\ \bibnamefont {Hawkins}},
  \bibinfo {author} {\bibfnamefont {M.~C.}\ \bibnamefont {Duff}}, \bibinfo
  {author} {\bibfnamefont {M.}~\bibnamefont {Groza}}, \bibinfo {author}
  {\bibfnamefont {V.}~\bibnamefont {Buliga}}, \ and\ \bibinfo {author}
  {\bibfnamefont {A.}~\bibnamefont {Burger}},\ }\href {\doibase
  10.1007/s11664-009-0763-x} {\bibfield  {journal} {\bibinfo  {journal} {J.
  Electron. Mater.}\ }\textbf {\bibinfo {volume} {38}},\ \bibinfo {pages}
  {1522} (\bibinfo {year} {2009})}\BibitemShut {NoStop}%
\bibitem [{\citenamefont {Morosan}\ \emph {et~al.}(2006)\citenamefont
  {Morosan}, \citenamefont {Zandbergen}, \citenamefont {Dennis}, \citenamefont
  {Bos}, \citenamefont {Onose}, \citenamefont {Klimczuk}, \citenamefont
  {Ramirez}, \citenamefont {Ong},\ and\ \citenamefont
  {Cava}}]{Morosan_2006_NaturePhys}%
  \BibitemOpen
  \bibfield  {author} {\bibinfo {author} {\bibfnamefont {E.}~\bibnamefont
  {Morosan}}, \bibinfo {author} {\bibfnamefont {H.~W.}\ \bibnamefont
  {Zandbergen}}, \bibinfo {author} {\bibfnamefont {B.~S.}\ \bibnamefont
  {Dennis}}, \bibinfo {author} {\bibfnamefont {J.~W.~G.}\ \bibnamefont {Bos}},
  \bibinfo {author} {\bibfnamefont {Y.}~\bibnamefont {Onose}}, \bibinfo
  {author} {\bibfnamefont {T.}~\bibnamefont {Klimczuk}}, \bibinfo {author}
  {\bibfnamefont {A.~P.}\ \bibnamefont {Ramirez}}, \bibinfo {author}
  {\bibfnamefont {N.~P.}\ \bibnamefont {Ong}}, \ and\ \bibinfo {author}
  {\bibfnamefont {R.~J.}\ \bibnamefont {Cava}},\ }\href
  {http://dx.doi.org/10.1038/nphys360} {\bibfield  {journal} {\bibinfo
  {journal} {Nature Phys.}\ }\textbf {\bibinfo {volume} {2}},\ \bibinfo {pages}
  {544} (\bibinfo {year} {2006})}\BibitemShut {NoStop}%
\bibitem [{\citenamefont {Morosan}\ \emph {et~al.}(2010)\citenamefont
  {Morosan}, \citenamefont {Wagner}, \citenamefont {Zhao}, \citenamefont {Hor},
  \citenamefont {Williams}, \citenamefont {Tao}, \citenamefont {Zhu},\ and\
  \citenamefont {Cava}}]{Morosan_2010_PRB}%
  \BibitemOpen
  \bibfield  {author} {\bibinfo {author} {\bibfnamefont {E.}~\bibnamefont
  {Morosan}}, \bibinfo {author} {\bibfnamefont {K.~E.}\ \bibnamefont {Wagner}},
  \bibinfo {author} {\bibfnamefont {L.~L.}\ \bibnamefont {Zhao}}, \bibinfo
  {author} {\bibfnamefont {Y.}~\bibnamefont {Hor}}, \bibinfo {author}
  {\bibfnamefont {A.~J.}\ \bibnamefont {Williams}}, \bibinfo {author}
  {\bibfnamefont {J.}~\bibnamefont {Tao}}, \bibinfo {author} {\bibfnamefont
  {Y.}~\bibnamefont {Zhu}}, \ and\ \bibinfo {author} {\bibfnamefont {R.~J.}\
  \bibnamefont {Cava}},\ }\href {\doibase 10.1103/PhysRevB.81.094524}
  {\bibfield  {journal} {\bibinfo  {journal} {Phys. Rev. B}\ }\textbf {\bibinfo
  {volume} {81}},\ \bibinfo {pages} {094524} (\bibinfo {year}
  {2010})}\BibitemShut {NoStop}%
\bibitem [{\citenamefont {Novello}\ \emph {et~al.}(2015)\citenamefont
  {Novello}, \citenamefont {Hildebrand}, \citenamefont {Scarfato},
  \citenamefont {Didiot}, \citenamefont {Monney}, \citenamefont {Ubaldini},
  \citenamefont {Berger}, \citenamefont {Bowler}, \citenamefont {Aebi},\ and\
  \citenamefont {Renner}}]{Novello_2015_PRB}%
  \BibitemOpen
  \bibfield  {author} {\bibinfo {author} {\bibfnamefont {A.~M.}\ \bibnamefont
  {Novello}}, \bibinfo {author} {\bibfnamefont {B.}~\bibnamefont {Hildebrand}},
  \bibinfo {author} {\bibfnamefont {A.}~\bibnamefont {Scarfato}}, \bibinfo
  {author} {\bibfnamefont {C.}~\bibnamefont {Didiot}}, \bibinfo {author}
  {\bibfnamefont {G.}~\bibnamefont {Monney}}, \bibinfo {author} {\bibfnamefont
  {A.}~\bibnamefont {Ubaldini}}, \bibinfo {author} {\bibfnamefont
  {H.}~\bibnamefont {Berger}}, \bibinfo {author} {\bibfnamefont {D.~R.}\
  \bibnamefont {Bowler}}, \bibinfo {author} {\bibfnamefont {P.}~\bibnamefont
  {Aebi}}, \ and\ \bibinfo {author} {\bibfnamefont {C.}~\bibnamefont
  {Renner}},\ }\href {\doibase 10.1103/PhysRevB.92.081101} {\bibfield
  {journal} {\bibinfo  {journal} {Phys. Rev. B}\ }\textbf {\bibinfo {volume}
  {92}},\ \bibinfo {pages} {081101(R)} (\bibinfo {year} {2015})}\BibitemShut
  {NoStop}%
\bibitem [{\citenamefont {Goli}\ \emph {et~al.}(2012)\citenamefont {Goli},
  \citenamefont {Khan}, \citenamefont {Wickramaratne}, \citenamefont {Lake},\
  and\ \citenamefont {Balandin}}]{Goli_2012_NanoLet}%
  \BibitemOpen
  \bibfield  {author} {\bibinfo {author} {\bibfnamefont {P.}~\bibnamefont
  {Goli}}, \bibinfo {author} {\bibfnamefont {J.}~\bibnamefont {Khan}}, \bibinfo
  {author} {\bibfnamefont {D.}~\bibnamefont {Wickramaratne}}, \bibinfo {author}
  {\bibfnamefont {R.~K.}\ \bibnamefont {Lake}}, \ and\ \bibinfo {author}
  {\bibfnamefont {A.~A.}\ \bibnamefont {Balandin}},\ }\href {\doibase
  10.1021/nl303365x} {\bibfield  {journal} {\bibinfo  {journal} {Nano Lett.}\
  }\textbf {\bibinfo {volume} {12}},\ \bibinfo {pages} {5941} (\bibinfo {year}
  {2012})}\BibitemShut {NoStop}%
\bibitem [{\citenamefont {Li}\ \emph {et~al.}(2016{\natexlab{b}})\citenamefont
  {Li}, \citenamefont {Zhao}, \citenamefont {Liu}, \citenamefont {Ren},
  \citenamefont {Eda},\ and\ \citenamefont {Loh}}]{Li_2016_APL}%
  \BibitemOpen
  \bibfield  {author} {\bibinfo {author} {\bibfnamefont {L.~J.}\ \bibnamefont
  {Li}}, \bibinfo {author} {\bibfnamefont {W.~J.}\ \bibnamefont {Zhao}},
  \bibinfo {author} {\bibfnamefont {B.}~\bibnamefont {Liu}}, \bibinfo {author}
  {\bibfnamefont {T.~H.}\ \bibnamefont {Ren}}, \bibinfo {author} {\bibfnamefont
  {G.}~\bibnamefont {Eda}}, \ and\ \bibinfo {author} {\bibfnamefont {K.~P.}\
  \bibnamefont {Loh}},\ }\href {\doibase 10.1063/1.4963885} {\bibfield
  {journal} {\bibinfo  {journal} {Appl. Phys. Lett.}\ }\textbf {\bibinfo
  {volume} {109}},\ \bibinfo {eid} {141902} (\bibinfo {year}
  {2016}{\natexlab{b}}),\ 10.1063/1.4963885}\BibitemShut {NoStop}%
\bibitem [{\citenamefont {Jiang}\ \emph {et~al.}(2016)\citenamefont {Jiang},
  \citenamefont {Kurouski}, \citenamefont {Pozzi}, \citenamefont {Chiang},
  \citenamefont {Hersam},\ and\ \citenamefont
  {Duyne}}]{Jiang_2016_ChemPhysLett}%
  \BibitemOpen
  \bibfield  {author} {\bibinfo {author} {\bibfnamefont {N.}~\bibnamefont
  {Jiang}}, \bibinfo {author} {\bibfnamefont {D.}~\bibnamefont {Kurouski}},
  \bibinfo {author} {\bibfnamefont {E.~A.}\ \bibnamefont {Pozzi}}, \bibinfo
  {author} {\bibfnamefont {N.}~\bibnamefont {Chiang}}, \bibinfo {author}
  {\bibfnamefont {M.~C.}\ \bibnamefont {Hersam}}, \ and\ \bibinfo {author}
  {\bibfnamefont {R.~P.~V.}\ \bibnamefont {Duyne}},\ }\href {\doibase
  http://dx.doi.org/10.1016/j.cplett.2016.06.035} {\bibfield  {journal}
  {\bibinfo  {journal} {Chem. Phys. Lett.}\ }\textbf {\bibinfo {volume}
  {659}},\ \bibinfo {pages} {16 } (\bibinfo {year} {2016})}\BibitemShut
  {NoStop}%
\end{thebibliography}
%

\end{document}